\documentclass[10pt]{article}

 \usepackage{amsmath,amsthm,amssymb}
 \usepackage{makeidx,epsfig,lscape}
 \usepackage{color,colortbl}
 \usepackage{fancyhdr}
 \usepackage{xcolor,pict2e}

 \renewcommand{\headrulewidth}{0pt}
 \renewcommand{\footrulewidth}{0.5pt}

 \definecolor{myaqua}{rgb}{0.0,0.5,0.55}
 \definecolor{lightaqua}{rgb}{0.75,0.95,0.95}

 \usepackage[colorlinks = true,
            linkcolor = myaqua,
            urlcolor  = blue,
            citecolor = myaqua]{hyperref}

\usepackage{caption}
\usepackage{floatrow}
\def\lin#1#2{\textcolor[rgb]{0.6,0.6,0.6}{\vspace*{#1mm} \hrule
   height 3 pt \vspace*{#2mm}}}
\def\bt{\begin{tabular}}
\def\et{\end{tabular}}
\def\and{\mbox{ and }}

\def\1{{\bf 1}}

 \def\sectionn#1{\refstepcounter{section}{\color{myaqua}

 \vskip 6mm

 \noindent\Large\bf\thesection. #1}

 \vskip 3mm}

 \def\boxx#1#2#3#4#5{
 {\linethickness{#4pt}\put(#1,#5){\color{myaqua}{\line(1,0){#3}}}}
 \multiput(#1,#2)(0,#4){2}{\line(1,0){#3}}
 \multiput(#1,#2)(#3,0){2}{\line(0,1){#4}}
  }

\begin{document}

 \fancyhead[L]{\hspace*{-13mm}
 \bt{l}{\bf International Journal of Astronomy \& Astrophysics, 2019, *,**}\\
 Published Online **** 2019 in SciRes.
 \href{http://www.scirp.org/journal/*****}{\color{blue}{\underline{\smash{http://www.scirp.org/journal/****}}}} \\
 \href{http://dx.doi.org/10.4236/****.2014.*****}{\color{blue}{\underline{\smash{http://dx.doi.org/10.4236/****.2014.*****}}}} \\
 \et}
 \fancyhead[R]{\includegraphics{pic1.ps}}

 $\mbox{ }$

 \vskip 12mm

{ 

{\noindent{\huge\bf\color{myaqua}
Spectroscopic Survey of H$\alpha$ Emission Line Stars Associated
with Bright Rimmed Clouds}}
%
\\[6mm]
{\large\bf Kensuke Hosoya$^1$, Yoichi Itoh$^1$, Yumiko Oasa$^2$,
           Ranjan Gupta$^3$, and A. K. Sen$^4$}}
\\[2mm]
{ 
 $^1$ Nishi-Harima Astronomical Observatory, Center for Astronomy, 
University of Hyogo, 407-2 Nishigaichi, Sayo, Sayo, Hyogo 679-5313, Japan\\
Email: \href{mailto:yitoh@nhao.jp}{\color{blue}{\underline{\smash{yitoh@nhao.jp}}}}\\[1mm]
$^2$Faculty of Education, Saitama University, 255 Shimo-Okubo, Sakura, 
Saitama, Saitama, Japan\\
$^3$Inter University Center for Astronomy and Astrophysics(IUCAA),
 Ganeshkhind, Pune 411 007, India\\
$^4$Department of Physics, Assam University, Silchar 788011, Assam, India\\
 \\[4mm]
Received **** 2019
 \\[4mm]
Copyright \copyright \ 2019 by author(s) and Scientific Research Publishing Inc. \\
This work is licensed under the Creative Commons Attribution International License (CC BY). \\
\href{http://creativecommons.org/licenses/by/4.0/}{\color{blue}{\underline{\smash{http://creativecommons.org/licenses/by/4.0/}}}}\\
 \includegraphics{pic2.ps}

\lin{5}{7}

 { 
 {\noindent{\large\bf\color{myaqua} Abstract}{\bf \\[3mm]
 \textup{
\textcolor{red}{The results of a} spectroscopic survey of
H$\alpha$ emission line stars associated with fourteen bright rimmed clouds
are presented.
Slit-less optical spectroscopy was carried out with the
Inter University Centre for Astronomy and Astrophysics (IUCAA) 
2m telescope and IUCAA Faint Object Spectrograph and Camera (IFOSC).
H$\alpha$ emission line was detected from 173 objects.
Among them 85 objects have a strong H$\alpha$ emission line
with its equivalent width larger than 10 \AA.
Those are classical T Tauri stars.
52 objects have a weak H$\alpha$ emission line with its equivalent width
less than 10 \AA~ and do not show intrinsic near-infrared excess.
Those are weak-line T Tauri stars.
On the other hand, 36 objects have a weak H$\alpha$ emission line ($<10$\AA),
although they show intrinsic near-infrared excess.
Such objects are not common in low-mass star forming regions.
Those are misfits of the general concept on formation process of a
low-mass star, in which it evolves from
a classical T Tauri star to a weak-line T Tauri star.
Those might be weak-line T Tauri stars with a flared disk 
in which gas is heated by ultraviolet radiation from a nearby early-type
star.
Alternatively, we propose pre-transitional disk objects as their evolutional stage.
 }}}
 \\[4mm]
 {\noindent{\large\bf\color{myaqua} Keywords}{\bf \\[3mm]
Star formation; Pre-main sequence stars; T Tauri stars
}

 \fancyfoot[L]{{\noindent{\color{myaqua}{\bf How to cite this
 paper:}} Hosoya et al. (2019)
Spectroscopic Survey of H$\alpha$ Emission Line Stars Associated
with Bright Rimmed Clouds.
 International Journal of Astronomy \& Astrophysics,*,***-***}}

\lin{3}{1}

\sectionn{Introduction}

{ \fontfamily{times}\selectfont
 \noindent 
A T Tauri star (TTS) is a low-mass star in the pre-main sequence stage.
It was first identified as a variable star \cite{Joy45}.
Successive studies on low-mass star forming regions,
such as the Taurus molecular cloud, revealed two types of TTSs,
namely classical TTSs (CTTSs) and weak-line TTSs (WTTSs).
CTTSs show the H$\alpha$ emission line with its equivalent width larger
than 10 \AA.
WTTSs also show the H$\alpha$ emission line, but its
equivalent width is less than 10 \AA.
\cite{Meyer97} investigated near-infrared colors
of TTSs in the Taurus molecular cloud.
They estimated the amount of the interstellar extinction from
$E(R-I)$ color and then calculated the dereddened colors of the TTSs.
They noted that the WTTSs have the dereddened colors similar to
those of normal main sequence stars.
On the other hand, the dereddened colors of CTTSs occupy
a narrow range in a $(J-H)-(H-K)$ color-color
diagram.
The range is well fitted by a linear line with an intersection with
the reddening vector from an M6 dwarf at $A_{\rm V}=1.9$ mag (hereafter termed
as the dereddened CTTS line).
Dereddened colors of approximately half of the CTTSs are redder than that 
intersection point.
Such CTTSs have an intrinsic near-infrared excess.
Those are brighter in the $H$- and/or $K$ band than the radiation expected 
only from the photosphere.
The intrinsic near-infrared excess of a TTS is caused by an optically
thick circumstellar disk, and the amount of the excess depends on the
optical thickness of the disk, geometric structure of the disk, and viewing
angle of the disk \cite{Bertout88}, \cite{Lada92}.
Approximately half of CTTSs show the intrinsic near-infrared excess while
the other CTTSs do not.
On the other hand, WTTSs do not show the intrinsic near-infrared excess.

A low-mass star evolves from a protostar to a CTTS and subsequently a WTTS.
\cite{Bertout07} investigated the ages of TTSs in the Taurus
molecular cloud.
They used Hipparcos parallaxes as well as photometric and spectroscopic 
information for 72 TTSs.
They plotted them on the HR diagram and then estimated their ages
by comparing them to a pre-main sequence evolutionary track.
It is revealed that the ages of the CTTSs are younger than those of the WTTSs.
A solar mass star evolves from a CTTS to a WTTS at the age of $4\times10^{6}$
yr.
\cite{Takagi14} deduced the surface gravity of
TTSs in the Taurus molecular cloud
from high resolution spectroscopy and then estimated their ages
by plotting them on the HR diagram.
The fraction of the objects with the intrinsic 
near-infrared excess decreases with the age.
They concluded that the dissipation timescale of the circumstellar 
disk is $3-4\times10^{6}$ yr.
The general concept that a low-mass star evolves from a CTTS to a WTTS
is well established by observational studies of nearby
low-mass star forming regions.

Bright rimmed clouds (BRCs) are molecular clouds harboring an IRAS source
and are located at periphery of an HII region.
Ultra-violet radiation from OB stars ionizes molecular cloud
materials through shock and excites hydrogen atoms at the surface
of the cloud, making diffuse H$\alpha$ emission.
\cite{Sugitani91} and \cite{Sugitani94}
presented comprehensive catalogs of BRCs, in which 89 BRCs are listed.
Properties of the clouds and young stellar objects (YSOs) associated
with the clouds have been extensively investigated.
\cite{Niwa09} carried out millimeter mapping observations of
molecular clouds associated with the W5 HII region.
They noticed that molecular clouds facing the HII region show a 
steep density gradient toward the HII region due to
the compression of the HII region to the clouds.
\cite{Ogura02} conducted H$\alpha$ grism spectroscopy
and optical narrow-band imaging observations of 28 BRCs.
They detected 460 H$\alpha$ emission line stars.
\cite{Hayashi12} carried out wide-field near-infrared
imaging observations of 32 BRCs.
They identified 2099 objects as YSO candidates.
These observational studies confirmed the formation of low-mass stars in BRCs.
\cite{Sugitani95} carried out a near-infrared survey
of 44 BRCs.
They indicated that stars with bluer colors are located closer to the OB
star and those with redder colors were closer to the IRAS source.
\cite{Matsuyanagi06} presented a near-infrared image of BRC 14.
They revealed that the fraction of YSO candidates to all sources, the extinction
of all sources, and the near-infrared excess of the YSO candidates increased
from the the outside of the rim to the center of the molecular cloud.
These results indicated that star formation proceeds
from outside to the center of the cloud.

We conducted slit-less optical spectroscopy of 14 BRCs to identify
H$\alpha$ emission line stars.
We discuss evolution process of a low-mass star in a massive-star
forming region based on the relationship between the H$\alpha$ emission line and
its intrinsic near-infrared excess.
}

\renewcommand{\headrulewidth}{0.5pt}
\renewcommand{\footrulewidth}{0pt}

 \pagestyle{fancy}
 \fancyfoot{}
 \fancyhead{} 
 \fancyhf{}
 \fancyhead[RO]{\leavevmode \put(-90,0){\color{myaqua}K. Hosoya, et al.} \boxx{15}{-10}{10}{50}{15} }
 \fancyhead[LE]{\leavevmode \put(0,0){\color{myaqua}K. Hosoya, et al.}  \boxx{-45}{-10}{10}{50}{15} }
 \fancyfoot[C]{\leavevmode
 \put(0,0){\color{lightaqua}\circle*{34}}
 \put(0,0){\color{myaqua}\circle{34}}
 \put(-2.5,-3){\color{myaqua}\thepage}}

 \renewcommand{\headrule}{\hbox to\headwidth{\color{myaqua}\leaders\hrule height \headrulewidth\hfill}}

\sectionn{Observations and Data Reduction}

{ \fontfamily{times}\selectfont
 \noindent
Spectroscopic survey of H$\alpha$ emission line stars were
carried out on six nights of
2011 January and 2012 January with IFOSC (IUCAA Faint Optical
Spectrograph and Camera) mounted on the IUCAA 2-m telescope at Giravali
near Pune, India.
We used the IFOSC5 grism and the wide H$\alpha$ filter, and we
did not use any slits.
IFOSC has a 2048 $\times$ 2048 CCD with the field of view of 
10.5' $\times$ 10.5'.
We had spectra centered at 6563 \AA~ with a width of 80 \AA~ and
spectral resolution of $\sim$ 9 \AA.
We also obtained $V$-band images.
The targets were 14 BRCs listed in \cite{Sugitani91}
and \cite{Sugitani94} 
that were observable on the observing date (Table \ref{tab:target}).
Three frames of 300-s exposure each were obtained for spectroscopy and 
one frame of 60 s was obtained for imaging.
The OB star making the HII region and the bright rim is
located in the observing fields of view for BRCs 15, 24, and 25.
However, these OB stars are so bright that their spectra were saturated.
The general seeing conditions varied
between 0.9'' and 2.0''.

\begin{table}
  \caption{Properties of the observed bright rimmed clouds}\label{tab:target}
  \begin{center}
    \begin{tabular}{ccccccc}
      \hline \hline
      Cloud & RA & DEC  & HII region & Spectral type & Distance & obs. fields\\ 
      & [J2000] & [J2000] & & of the exciting star & [pc] & \\ 
      \hline
      BRC 15 & 05 23 30.1 & +33 11 54 & S 236  & O5V,O6V & 3600 & 6\\
      BRC 16 & 05 19 48.9 & -05 52 05 & S 276  & O7V,O8.5II,O9.5I & 400 & 2\\
      BRC 17 & 05 31 28.1 & +12 06 24 & S 264  & O5V & 450 & 3\\
      BRC 18 & 05 44 29.8 & +09 08 54 & S 264  & O5V & 450 & 4\\
      BRC 19 & 05 34 30.7 & -02 38 12 & S 277  & O9.5V & 400 & 1\\
      BRC 21 & 05 39 41.3 & -03 37 12 & S 277  & O9.5V & 400 & 1\\
      BRC 23 & 06 22 58.7 & +23 09 58 & S 249  & O9V &1900 & 1\\
      BRC 24 & 06 34 55.4 & +04 25 14 & S 275  & O5V &1400 & 9\\
      BRC 25 & 06 41 03.3 & +10 21 59 & S 273  & O7V & 913 & 3\\
      BRC 26 & 07 03 47.2 & -11 45 47 & S 296  & O7III & 990 & 2\\
      BRC 27 & 07 03 58.7 & -11 23 19 & S 296  & B0V & 990 & 2\\
      BRC 28 & 07 04 43.4 & -10 21 59 & S 296  & B0IV & 990 & 2\\
      BRC 29 & 07 04 52.4 & -12 09 26 & S 296  & O7.5V & 990 & 1\\
      BRC 45 & 07 18 23.7 & -22 06 13 & RCW 14 & O8V & 1930 & 2\\
      \hline 
    \end{tabular}
  \end{center}
\end{table}

The object frames were calibrated with the Image Reduction and Analysis Facility
(IRAF). 
Data were processed in standard manner, namely bias
subtraction and flat fielding with the twilight frames.
We detected point sources in the $V$-band image with SExtractor program.
The limiting magnitude is approximately 19 mag.
Based on the coordinate of the source, a spectrum image of each object was
extracted from the spectral frame.
The spectrum extends along a line.
We average the counts of each line and
the average count was then subtracted from each line of the image.
In the process, the continuum flux of the object was subtracted 
and the emission line appeared as a point source.
We detected the emission line via the SExtractor program.
We also confirmed the emission line by eye inspection.
For the image with the emission line, we extracted a 1-D spectrum
from the image prior to the continuum subtraction.
An equivalent width of the emission line was measured with the IRAF splot
task.
We did not fit a Gaussian profile.
The minimum equivalent width of the detected H$\alpha$ emission line
was 0.3 \AA.
}

\sectionn{Results}
{ \fontfamily{times}\selectfont
 \noindent
Emission lines of H$\alpha$ were identified from 173 objects (Table \ref{tab:object}).
Their equivalent widths range from 0.3 \AA~ to 132 \AA.
We investigated near-infrared properties of the emission line stars
by using 2MASS photometries.
The near-infrared color-color diagram of the emission line stars are
presented in Figure \ref{fig:BRCcc}.
We defined the line parallel to the reddening vector through an M6 dwarf color
as the near-infrared excess border.
Among the emission line stars, 77 objects are plotted redward of the border.
We identified that such objects have the intrinsic near-infrared excess.
The color-color diagram of TTSs in the BRCs is significantly different from 
that of TTSs in Taurus.
TTSs with the equivalent width less than 10 \AA~ in Taurus are plotted blueward 
of the near-infrared excess
border (Meyer et al. 1997).
On the other hand, such objects in the BRCs are plotted on either side of 
the near-infrared excess border,
like CTTSs in the BRCs and also like CTTSs in Taurus.

In the Taurus molecular cloud, WTTSs are older than CTTSs.
The ages of the TTSs in the BRCs were estimated on the ($I$, $I-J$) 
color-magnitude diagram
with the isochrone of \cite{Siess00}.
$I$-magnitudes of the objects were taken from the USNO-B1.0 catalog.
Because the extinction vector is relatively parallel to the isochrone
on this diagram, it is possible to roughly estimate the age of the object.
It is indicated that majority of the objects have the age between 1 Myr 
and 10 Myr.
We did not find any differences in the
ages between the objects with a strong H$\alpha$
emission line and the objects with a weak H$\alpha$ emission line.

Masses of the TTSs were estimated with the isochrones of \cite{Siess00},
\cite{Baraffe98}, and \cite{Baraffe03} on the ($J$, $J-H$) color-magnitude
diagram.
It is revealed that most of the objects have mass between 0.5 M$_{\odot}$
and 2 M$_{\odot}$.

\begin{table}
  \caption{H$\alpha$ emission line stars}\label{tab:object}
  \begin{center}
    \begin{tabular}{ccccllll}
      \hline \hline
      ID & RA [J2000] & DEC [J2000] & EW(H$\alpha$) & $J$ & $H$ & $K$ & Identification \\ 
      & $^{h~~~~~m~~~~~s}$ & $^{\circ~~~~~'~~~~~"}$ & \AA & mag & mag & mag \\ 
      \hline
\multicolumn{7}{c}{BRC 15} \\
      \hline
1  & 5 23 32.21 &  33 27 43.6 &  16.8 & 14.36 & 13.29 & 12.41 &[PSM2011] 107\\
2  & 5 23 15.81 &  33 27 43.5 &   3.8 & 15.17 & 14.25 & 13.96 &[PSM2011] 135, [CMP2012] 975\\
3  & 5 22 54.07 &  33 26 33.0 &   1.2 & 14.35 & 13.55 & 13.25 &[PSM2011]  82, [CMP2012] 559\\
4  & 5 22 51.19 &  33 26 32.9 &  11.4 & 15.96 & 15.19 & 14.59 &[MSB2007]37, [PSM2011] 851, [CMP2012] 465\\
5  & 5 22 53.82 &  33 25 41.5 &  13.9 & 15.93 & 14.89 & 14.12 &[MSB2007]38, [PSM2011] 883, [CMP2012] 548\\
6  & 5 22 52.30 &  33 24  7.4 &   1.7 & 13.58 & 12.71 & 12.25 &[PSM2011]  61, [CMP2012] 496\\
7  & 5 22 51.90 &  33 23 59.3 &   1.1 & 14.61 & 14.21 & 13.65 &[PSM2011] 56\\
8  & 5 22 51.04 &  33 25 47.1 &   2.5 & 13.08 & 12.34 & 11.75 &[MSB2007]41, [PSM2011] 847, [CMP2012] 461\\
9  & 5 22 54.15 &  33 24 58.0 &   2.7 & 15.15 & 14.46 & 14.20 &[PSM2011] 1168,[CMP2012] 563\\
10 & 5 22 45.78 &  33 28 16.2 &  60   & 15.05 & 14.04 & 13.54 &[PSM2011] 177, [CMP2012] 294\\
11 & 5 22 18.44 &  33 28 21.2 &   6.3 & 11.13 & 11.04 & 10.98 &NGC1893-256, [PSM2011] 1072,[CMP2012] 11\\
12 & 5 22 51.04 &  33 25 47.1 &   1.1 & 13.08 & 12.34 & 11.75 &[MSB2007]41, [PSM2011] 847, [CMP2012] 461\\
13 & 5 22 43.02 &  33 25  5.4 &  24.5 & 11.65 & 11.31 & 10.92 &NGC1893-35, [MSB2007]47, [PSM2011]  12 \\
14 & 5 22 43.78 &  33 25 25.8 &  40.1 & 12.38 & 10.77 &  9.42 &[MSB2007]46, [PSM2011]  92, [CMP2012] 241\\
15 & 5 22 38.78 &  33 22  5.4 &   8.6 & 14.68 & 13.70 & 13.08 &[PSM2011] 718, [CMP2012] 151\\
16 & 5 22 46.84 &  33 29 27.9 &   9.6 & 14.83 & 13.9  & 13.36 &[MSB2007]40, [PSM2011] 131, [CMP2012] 329\\
17 & 5 22 49.35 &  33 29  2.1 &   7.1 & 14.99 & 14.06 & 13.57 &[PSM2011] 112, [CMP2012] 410\\
18 & 5 22 49.57 &  33 30  1.5 &  28.8 & 14.42 & 13.45 & 12.95 &[PSM2011] 137, [CMP2012] 414\\
19 & 5 23  9.33 &  33 30  2.3 &  35.7 & 10.79 & 10.41 & 10.01 &GGA 333, [PSM2011] 5\\
20 & 5 23  6.32 &  33 31  1.7 &   6.4 & 14.74 & 13.82 & 13.24 &[PSM2011]  97, [CMP2012] 873\\
21 & 5 23  4.96 &  33 31 50.1 &   1.5 & 15.14 & 14.38 & 13.89 &[PSM2011] 105, [CMP2012] 848\\
22 & 5 22 52.23 &  33 29 58.0 &   6.8 & 12.68 & 11.91 & 11.35 &[MSB2007]34, [PSM2011]  31, [CMP2012] 494\\
23 & 5 23  3.32 &  33 29 25.1 &   1.7 & 15.46 & 14.41 & 13.72 &[MSB2007]13, [PSM2011] 113, [CMP2012] 806\\
24 & 5 23  2.76 &  33 29 40.2 &   8.8 & 15.47 & 14.60 & 13.90 &[MSB2007]17, [PSM2011] 175, [CMP2012] 795\\
25 & 5 23  0.07 &  33 30 39.0 &  21.7 & 14.34 & 13.42 & 12.97 &[PSM2011]  99, [CMP2012] 718\\
26 & 5 22 58.11 &  33 30 41.0 &   9.3 & 13.84 & 13.30 & 12.63 &[MSB2007]27, [PSM2011] 33\\
27 & 5 23 13.59 &  33 29 44.0 &   5.2 & 15.34 & 14.48 & 13.93 &[PSM2011] 144, [CMP2012] 961\\
28 & 5 23  1.07 &  33 29 25.1 &  30   & 15.97 & 14.89 & 14.19 &[MSB2007]21, [PSM2011] 253, [CMP2012] 743\\
29 & 5 23  9.95 &  33 29  8.8 &  12.6 & 15.40 & 14.11 & 13.20 &[MSB2007] 2, [PSM2011] 140, [CMP2012] 924\\
19 & 5 23  9.33 &  33 30  2.3 &  35.7 & 10.79 & 10.41 & 10.01 &GGA 333, [PSM2011] 5\\
20 & 5 23  6.32 &  33 31  1.7 &   6.4 & 14.74 & 13.82 & 13.24 &[PSM2011]  97, [CMP2012] 873\\
21 & 5 23  4.96 &  33 31 50.1 &   1.5 & 15.14 & 14.38 & 13.89 &[PSM2011] 105, [CMP2012] 848\\
22 & 5 22 52.23 &  33 29 58.0 &   6.8 & 12.68 & 11.91 & 11.35 &[MSB2007]34, [PSM2011]  31, [CMP2012] 494\\
23 & 5 23  3.32 &  33 29 25.1 &   1.7 & 15.46 & 14.41 & 13.72 &[MSB2007]13, [PSM2011] 113, [CMP2012] 806\\
24 & 5 23  2.76 &  33 29 40.2 &   8.8 & 15.47 & 14.60 & 13.90 &[MSB2007]17, [PSM2011] 175, [CMP2012] 795\\
25 & 5 23  0.07 &  33 30 39.0 &  21.7 & 14.34 & 13.42 & 12.97 &[PSM2011]  99, [CMP2012] 718\\
26 & 5 22 58.11 &  33 30 41.0 &   9.3 & 13.84 & 13.30 & 12.63 &[MSB2007]27, [PSM2011] 33\\
27 & 5 23 13.59 &  33 29 44.0 &   5.2 & 15.34 & 14.48 & 13.93 &[PSM2011] 144, [CMP2012] 961\\
28 & 5 23  1.07 &  33 29 25.1 &  30   & 15.97 & 14.89 & 14.19 &[MSB2007]21, [PSM2011] 253, [CMP2012] 743\\
29 & 5 23  9.95 &  33 29  8.8 &  12.6 & 15.40 & 14.11 & 13.20 &[MSB2007] 2, [PSM2011] 140, [CMP2012] 924\\
30 & 5 23  3.82 &  33 29 24.2 &   2   & 14.80 & 13.90 & 13.30 &[MSB2007]12, [PSM2011] 114, [CMP2012] 819\\
31 & 5 23  4.05 &  33 29 48.4 &  26.3 & 15.03 & 13.95 & 13.15 &[MSB2007]16, [PSM2011] 201, [CMP2012] 825\\
      \hline 
    \end{tabular}
  \end{center}
\end{table}

\begin{table}
  \begin{center}
    \begin{tabular}{ccccllll}
      \hline \hline
      ID & RA [J2000] & DEC [J2000] & EW(H$\alpha$) & $J$ mag & $H$ mag & $K$ mag & Identification\\ 
      & $^{h~~~~~m~~~~~s}$ & $^{\circ~~~~~'~~~~~"}$ & \AA & mag & mag & mag \\ 
      \hline
\multicolumn{7}{c}{BRC 17} \\
      \hline
1  & 5 31 23.12 &  12 10 33.7 &   6.9 & 13.17 & 12.46 & 12.02 \\
2  & 5 31 26.94 &  12 10 20.5 &   2.7 & 11.96 & 11.25 & 10.89 \\
3  & 5 31 23.59 &  12  9 43.9 &   5.2 & 10.42 &  9.62 &  9.19 & HI Ori, HBC 93, [DM99]143\\
4  & 5 31 28.05 &  12  9 10.3 &  16.8 &  9.41 &  8.31 &  7.34 & HK Ori, HBC 94\\

5  & 5 31 15.51 &  12 11 23.7 &   2.5 & 12.10 & 11.21 & 10.72 &[DM99]136\\
6  & 5 31 19.46 &  12  9 15.3 &   2.2 & 11.99 & 10.78 & 10.09 \\
7  & 5 30 51.70 &  12  8 36.7 &  26.5 & 11.64 & 10.87 & 10.45 &V4480 Ori, HBC 91, [DM99]125\\
8  & 5 31 12.49 &  12  7 54.5 &   2   & 13.29 & 12.52 & 12.21 \\
9  & 5 31 21.91 &  11 54 56.7 &   2.2 & 13.60 & 12.96 & 12.69 \\
      \hline
\multicolumn{7}{c}{BRC 18} \\
      \hline
1  & 5 44 19.40 &   9 16 19.3 &   3.5 & 13.12 & 12.39 & 12.03 \\
2  & 5 44 22.89 &   9 10 35.3 &   2.5 & 12.45 & 11.72 & 11.45 &[DM99]247\\
3  & 5 44 23.21 &   9 12  3.9 &   1.4 & 10.00 &  9.36 &  8.95 &[DM99]248\\
4  & 5 44 37.32 &   9 11 59.1 &   3.8 & 12.95 & 12.33 & 12.05 \\
5  & 5 44 37.02 &   9 13 20.0 &  15.4 & 12.33 & 11.66 & 11.33 &[DM99]252\\
6  & 5 44 30.23 &   9 12 23.8 &   8.1 & 14.13 & 13.36 & 12.91 \\
7  & 5 44 17.35 &   9 10 59.2 &  17   & 10.79 & 10.04 &  9.62 &V630 Ori, [DM99]244\\
8  & 5 44 25.73 &   9 12  1.3 &   1.9 & 13.07 & 12.35 & 12.09 \\
9  & 5 44 10.04 &   9  8 40.8 &  43.2 & 13.85 & 13.12 & 12.64 \\
10 & 5 44 14.86 &   9  8  8.2 &   9.4 & 12.67 & 11.97 & 11.67 &[DM99]243\\
11 & 5 44  7.26 &   9  6 38.1 &  41.9 & 11.66 & 10.94 & 10.68 &V629 Ori, [DM99]240\\
12 & 5 44 28.66 &   9  6 18.6 &   8.8 & 14.43 & 13.81 & 13.46 \\
13 & 5 44  8.99 &   9  9 14.8 &   3.4 & 11.10 & 10.03 &  9.25 &QR Ori, [DM99]241\\
14 & 5 44 12.77 &   9  5 10.5 &  10.5 & 13.26 & 12.59 & 12.24 \\
15 & 5 43 56.59 &   9 18 16.9 &  26.8 & 12.85 & 12.04 & 11.67 &[DM99]238\\
16 & 5 43 31.22 &   9 17 46.3 &  10.8 & 12.88 & 12.30 & 11.97 \\
17 & 5 43 57.00 &   9 16 24.0 &  38.8 & 12.34 & 11.57 & 11.17 \\
18 & 5 43 50.67 &   9 12 25.0 &  10.3 & 13.33 & 12.37 & 11.82 \\
19 & 5 43 39.59 &   9 10 40.8 &   4.9 & 13.34 & 12.66 & 12.35 \\
20 & 5 43 51.36 &   9 13 40.6 &  66.9 & 14.62 & 13.98 & 13.74 \\
21 & 5 43 32.20 &   9  9 24.0 &   7.2 & 11.29 & 10.55 & 10.19 &V626 Ori, [DM99]231\\
22 & 5 43 44.50 &   9  8  1.2 &  20.8 & 14.57 & 12.62 & 11.09 \\
23 & 5 43 53.50 &   9  7 11.8 &  42   & 13.09 & 12.23 & 11.71 \\
24 & 5 43 20.92 &   9  6  7.1 &   5   & 10.07 &  9.13 &  8.41 &V625 Ori, [DM99]227\\
25 & 5 43 52.20 &   9  6 51.4 &   1.8 & 14.45 & 13.82 & 13.56 \\
      \hline
\multicolumn{7}{c}{BRC 19} \\
      \hline
1  & 5 34 34.24 &  -2 58 16.6 &  55   & 12.77 & 11.67 & 11.03 & Haro 5-83\\
2  & 5 34 20.38 &  -2 57 46.9 &   3.1 & 13.75 & 12.27 & 11.27 &V1945 Ori\\
      \hline 
    \end{tabular}
  \end{center}
\end{table}

\begin{table}
  \begin{center}
    \begin{tabular}{ccccllll}
      \hline \hline
      ID & RA [J2000] & DEC [J2000] & EW(H$\alpha$) & $J$ mag & $H$ mag & $K$ mag & Identification\\ 
      & $^{h~~~~~m~~~~~s}$ & $^{\circ~~~~~'~~~~~"}$ & \AA & mag & mag & mag \\ 
      \hline
\multicolumn{7}{c}{BRC 24} \\
      \hline
1  & 6 34 25.74 &   4 28 15.8 &  20.6 & 14.57 & 13.55 & 12.76 &[WFT2009] RMCX 324\\
2  & 6 34 41.04 &   4 27  9.9 &  15.9 & 14.82 & 13.50 & 12.68 \\
3  & 6 33 50.40 &   4 32 55.9 &   3.5 & 13.11 & 12.42 & 11.98 \\
4  & 6 34  2.21 &   4 30  7.6 &   9.9 & 14.43 & 13.49 & 13.00 &[WFT2009] RMCX 189\\
5  & 6 34 17.07 &   4 27 35.3 &  22.5 & 14.72 & 13.53 & 12.69 \\
6  & 6 34  4.22 &   4 34 34.1 &   9.9 & 14.45 & 13.44 & 12.86 &[WFT2009] RMCX 199\\
7  & 6 34  3.41 &   4 34  8.8 &  22   & 13.24 & 12.36 & 11.72 &[WFT2009] RMCX 195\\
8  & 6 33 58.01 &   4 33 31.3 &  20.2 & 14.75 & 13.71 & 12.97 &[WFT2009] RMCX 179\\
9  & 6 33 55.73 &   4 28 23.4 &  14.4 & 15.48 & 14.41 & 13.85 \\
10 & 6 33 54.60 &   4 29 41.6 &   6   & 15.35 & 13.87 & 12.88 \\
11 & 6 34  8.90 &   4 29 38.5 &  13.7 & 15.15 & 14.13 & 13.64 &[WFT2009] RMCX 233\\
12 & 6 33 45.75 &   4 31 3.5  &  25.4 & 15.24 & 14.08 & 13.56 \\
13 & 6 33 56.09 &   4 31 19.3 &  32.1 & 14.65 & 13.65 & 13.11 &[WFT2009] RMCX 172\\
14 & 6 34  2.85 &   4 34 51.0 &   2.4 & 13.55 & 12.68 & 12.20 &[WFT2009] RMCX 190\\
15 & 6 33 41.16 &   4 37  5.1 &   3.8 & 14.01 & 12.94 & 12.17 &[WFT2009] RMCX 149\\
16 & 6 32 35.75 &   4 46 30.4 &  23   & 14.01 & 13.10 & 12.56 & [BNM2013] 66.01 15\\
4  & 6 34  2.21 &   4 30  7.6 &   9.9 & 14.43 & 13.49 & 13.00 &[WFT2009] RMCX 189\\
5  & 6 34 17.07 &   4 27 35.3 &  22.5 & 14.72 & 13.53 & 12.69 \\
6  & 6 34  4.22 &   4 34 34.1 &   9.9 & 14.45 & 13.44 & 12.86 &[WFT2009] RMCX 199\\
7  & 6 34  3.41 &   4 34  8.8 &  22   & 13.24 & 12.36 & 11.72 &[WFT2009] RMCX 195\\
8  & 6 33 58.01 &   4 33 31.3 &  20.2 & 14.75 & 13.71 & 12.97 &[WFT2009] RMCX 179\\
9  & 6 33 55.73 &   4 28 23.4 &  14.4 & 15.48 & 14.41 & 13.85 \\
10 & 6 33 54.60 &   4 29 41.6 &   6   & 15.35 & 13.87 & 12.88 \\
11 & 6 34  8.90 &   4 29 38.5 &  13.7 & 15.15 & 14.13 & 13.64 &[WFT2009] RMCX 233\\
12 & 6 33 45.75 &   4 31 3.5  &  25.4 & 15.24 & 14.08 & 13.56 \\
13 & 6 33 56.09 &   4 31 19.3 &  32.1 & 14.65 & 13.65 & 13.11 &[WFT2009] RMCX 172\\
14 & 6 34  2.85 &   4 34 51.0 &   2.4 & 13.55 & 12.68 & 12.20 &[WFT2009] RMCX 190\\
15 & 6 33 41.16 &   4 37  5.1 &   3.8 & 14.01 & 12.94 & 12.17 &[WFT2009] RMCX 149\\
16 & 6 32 35.75 &   4 46 30.4 &  23   & 14.01 & 13.10 & 12.56 & [BNM2013] 66.01 15\\
17 & 6 32 34.95 &   4 44 39.2 &  32.8 & 10.11 &  9.90 &  9.53 & NGC2244 279\\
18 & 6 32 45.11 &   4 45 23.1 &  15.5 & 13.33 & 12.29 & 11.57 & [WTF2008] Main 878\\
19 & 6 32 31.01 &   4 50  6.0 &   1.5 & 12.07 & 11.57 & 11.09 & NGC2244 269\\
20 & 6 32 31.44 &   4 42 34.0 &   0.5 & 11.74 & 11.03 & 10.83 & [WTF2008] Main 831\\
21 & 6 32 50.74 &   4 44 47.6 &   1.3 & 12.94 & 12.09 & 11.70 & [WTF2008] Main 898\\
22 & 6 32  1.83 &   4 53 38.6 &   6.5 & 14.26 & 13.32 & 12.91 & [WTF2008] Main 498\\
23 & 6 32  7.82 &   4 52 28.4 &  18.3 & 14.03 & 13.21 & 12.76 & [WTF2008] Main 590\\
24 & 6 32  4.66 &   4 54 51.5 &   2.8 & 13.80 & 12.92 & 12.54 & [WTF2008] Main 541\\
25 & 6 32  4.53 &   4 53 25.0 &   2.7 & 13.59 & 12.71 & 12.23 & [WTF2008] Main 536\\
26 & 6 31 41.02 &   4 54 47.9 &   8.3 & 14.18 & 13.36 & 12.95 & [WTF2008] Main 133\\
27 & 6 31 43.85 &   5  2 57.5 &   4   & 13.27 & 12.38 & 11.97 & [WTF2008] Main 169\\
28 & 6 31 49.26 &   4 57  0.9 &   5.4 & 13.7  & 12.85 & 12.27 & [WTF2008] Main 254\\
29 & 6 31 51.65 &   4 55  5.2 &   6.7 & 13.91 & 13.01 & 12.45 & V547 Mon\\
30 & 6 31 29.76 &   4 54 49.1 &   8.8 & 11.45 & 10.89 &  9.69 & GGA 395, NGC2244 74\\
31 & 6 31 39.62 &   4 59 45.1 &  23.5 & 14.12 & 13.13 & 12.65 & [WTF2008] Main 124\\
32 & 6 31 39.88 &   4 56 39.1 &  17.8 & 15.05 & 14.10 & 13.62 \\
33 & 6 31 20.87 &   5  4  8.5 &   3.8 & 14.07 & 13.23 & 12.71 & [BNM2013] 66.03 296\\
34 & 6 31 48.31 &   4 58 20.3 &   5.8 & 12.43 & 11.57 & 11.08 & [WTF2008] Main 243\\
35 & 6 31 29.52 &   4 54 34.2 &   8.1 & 14.92 & 14.11 & 13.72 & [WTF2008] Main 43\\
36 & 6 30 42.71 &   4 55 31.4 &   1.7 & 11.23 & 10.47 &  9.75 & V539 Mon\\
      \hline 
    \end{tabular}
  \end{center}
\end{table}

\begin{table}
  \begin{center}
    \begin{tabular}{ccccllll}
      \hline \hline
      ID & RA [J2000] & DEC [J2000] & EW(H$\alpha$) & $J$ mag & $H$ mag & $K$ mag & Identification\\ 
      & $^{h~~~~~m~~~~~s}$ & $^{\circ~~~~~'~~~~~"}$ & \AA & mag & mag & mag \\ 
      \hline
37 & 6 31  0.55 &   4 58  7.3 &   8.4 & 14.64 & 13.76 & 13.18 & [WFT2010] 126\\
38 & 6 31 30.96 &   5  6 58.5 &  28.6 & 12.88 & 12.08 & 11.60 \\
39 & 6 31 40.00 &   5  5 56.4 &   6.3 &  9.68 &  8.96 &  8.27 & NGC2244 106\\
40 & 6 31  2.88 &   5  3 49.3 &   8.9 & 13.74 & 12.84 & 12.38 & [WFT2010] 134\\
41 & 6 31 12.63 &   5  5  2.6 &   3.6 & 14.42 & 13.30 & 12.62 \\
      \hline
\multicolumn{7}{c}{BRC 25} \\
      \hline
1  & 6 41  1.53 &  10 14 56.1 &  20.7 & 12.93 & 12.18 & 11.66 & LkH$\alpha$ 46\\

2  & 6 41  7.48 &  10 15  4.5 &   3.3 & 13.24 & 12.47 & 12.15 & ESO-H$\alpha$ 493\\
3  & 6 41  9.85 &  10 15  2.6 &  11.1 & 13.19 & 11.80 & 10.85 & ESO-H$\alpha$ 504\\
4  & 6 41  7.01 &  10 16 28.9 &   1.3 & 12.99 & 12.32 & 12.06 \\
5  & 6 40 59.79 &  10  2 12.6 &  15.5 & 14.66 & 13.82 & 13.50 & ESO-H$\alpha$ 454\\
6  & 6 41  2.86 &  10  7 11.9 &   4.5 & 13.98 & 13.22 & 12.87 & ESO-H$\alpha$ 468\\
7  & 6 41  7.26 &   9 58 31.2 &   0.3 & 12.54 & 11.87 & 11.68 & V609 Mon\\
8  & 6 40 49.21 &   9 57 38.8 &  12.4 & 13.57 & 12.74 & 12.37 & ESO-H$\alpha$ 415\\
9  & 6 41 17.25 &   9 54 32.4 &  25.6 & 12.83 & 12.05 & 11.55 & IQ Mon\\
10 & 6 40 49.75 &   9 52 58.6 &   2.7 & 14.20 & 13.56 & 13.33 & V343 Mon\\
11 & 6 40 50.59 &   9 54 57.3 &   6.1 & 12.47 & 11.69 & 11.28 & LkH$\alpha$ 31\\

12 & 6 40 52.55 &   9 52  6.0 &   2   & 12.69 & 11.97 & 11.72 & ESO-H$\alpha$ 425\\
13 & 6 41 12.58 &   9 52 31.2 &   9.6 & 11.49 & 10.69 & 10.21 & MM Mon\\
14 & 6 40 59.45 &   9 59 45.5 &   4.9 & 12.78 & 12.09 & 11.69 & V602 Mon\\
15 & 6 40 54.88 &   9 53 12.4 &   5.5 & 13.92 & 13.09 & 12.40 & ESO-H$\alpha$ 435\\
16 & 6 40 46.95 &   9 52 40.6 &  42   & 14.75 & 14.06 & 13.41 & ESO-H$\alpha$ 405\\
17 & 6 40 54.11 &   9 52 24.6 &  28.9 & 15.34 & 14.88 & 14.33 & ESO-H$\alpha$ 432\\
18 & 6 40 54.20 &   9 55 52.0 &  32   & 13.70 & 12.94 & 12.76 & NGC2264 SBL 219\\ 
      \hline
\multicolumn{7}{c}{BRC 26} \\
      \hline
1  & 7  3 40.75 & -11 46 16.6 &  22.7 & 15.24 & 14.06 & 13.21 \\
2  & 7  3 48.22 & -11 41 46.0 &   5.2 & 13.61 & 12.84 & 12.55 \\
3  & 7  4  7.18 & -11 40 32.7 &   3.9 & 14.08 & 13.87 & 13.73 \\
4  & 7  3 49.91 & -11 31 17.0 &   8.2 & 14.03 & 13.20 & 12.95 \\
5  & 7  3 54.00 & -11 32 47.9 &   5.2 & 12.34 & 11.61 & 11.18 \\
6  & 7  3 54.03 & -11 32 37.2 &  63.1 & 14.57 & 13.7  & 13.29 \\
7  & 7  3 58.35 & -11 33 34.6 &  19.1 & 14.92 & 13.93 & 13.52 \\
8  & 7  4  1.33 & -11 36 17.5 &  26.3 & 14.77 & 13.93 & 13.42 \\
9  & 7  4  0.68 & -11 37 49.3 &  12.8 & 13.98 & 13.23 & 12.93 \\
10 & 7  3 38.52 & -11 31 51.9 &  22.8 & 14.22 & 13.28 & 12.75 \\
11 & 7  3 43.68 & -11 36 56.6 &  33.3 & 14.89 & 13.92 & 13.66 \\
12 & 7  4  4.70 & -11 31 15.9 &  24.5 & 15.57 & 14.66 & 14.38 \\
13 & 7  3 58.06 & -11 32 39.9 &   3.4 & 14.10 & 13.23 & 12.93 \\
14 & 7  3 50.34 & -11 32 51.4 &  16.3 & 15.81 & 14.92 & 14.30 & [CPD2009] J070350.34-113251.4\\
15 & 7  4  0.41 & -11 33 59.6 &   2.4 & 12.10 & 10.92 & 10.19 & [CPD2009] J070400.41-113359.5\\
16 & 7  3 56.71 & -11 35  9.6 &  23.3 & 14.34 & 12.95 & 12.11 & [CPD2009] J070356.71-113509.6\\
17 & 7  4 3.94  & -11 35 57.1 &  11.6 & 14.51 & 13.63 & 13.21 \\
18 & 7  4  5.53 & -11 39 52.4 &  28.2 & 15.62 & 14.72 & 14.20 \\
19 & 7  3 48.98 & -11 37 19.4 &  20   & 15.42 & 14.58 & 14.24 \\
      \hline 
    \end{tabular}
  \end{center}
\end{table}

\begin{table}
  \begin{center}
    \begin{tabular}{ccccllll}
      \hline \hline
      ID & RA [J2000] & DEC [J2000] & EW(H$\alpha$) & $J$ mag & $H$ mag & $K$ mag & Identification\\ 
      & $^{h~~~~~m~~~~~s}$ & $^{\circ~~~~~'~~~~~"}$ & \AA & mag & mag & mag \\ 
      \hline
\multicolumn{7}{c}{BRC 27} \\
      \hline
1  & 7  4  8.04 & -11 23 54.7 &  16.9 & 13.12 & 12.44 & 12.20 & [OSP2002] BRC 27 22\\
2  & 7  4  6.70 & -11 26  8.5 &  17   & 10.76 & 10.02 &  9.12 & LkH$\alpha$ 220\\
3  & 7  4  9.96 & -11 23 16.4 &  18   & 11.79 & 10.71 &  9.83 & [OSP2002] BRC 27 25\\
4  & 7  3 53.04 & -11 29 35.3 &  10.9 & 12.36 & 11.54 & 10.95 & LkH$\alpha$ 219\\
5  & 7  3 54.03 & -11 32 37.2 &  88.3 & 14.57 & 13.7  & 13.29 \\
6  & 7  3 49.9  & -11 31 17.0 &  10.1 & 14.03 & 13.20 & 12.95 \\
7  & 7  3 45.76 & -11 23 14.9 &   5.8 & 13.59 & 12.59 & 12.19 \\
8  & 7  4  4.27 & -11 23 55.7 & 100   & 15.01 & 14.02 & 13.54 & [OSP2002] BRC 27 12\\
9  & 7  4  5.93 & -11 23 58.7 &   6.1 & 14.42 & 13.47 & 12.93 & [OSP2002] BRC 27 16\\
10 & 7  4 12.93 & -11 24  3.2 &   5.9 & 15.37 & 14.39 & 13.91 & [OSP2002] BRC 27 28\\
11 & 7  3 47.58 & -11 22 33.9 &   3.5 & 13.93 & 13.15 & 12.85 \\
12 & 7  3 57.05 & -11 30 16.9 & 131.8 & 15.30 & 14.63 & 14.30 \\
13 & 7  4 22.75 & -11 27 50.9 &  12.8 & 14.24 & 13.36 & 12.95 \\
14 & 7  4 20.87 & -11 29 36.2 &  18.6 & 13.80 & 12.88 & 12.29 & [CPD2009] J070420.87-112936.2\\
15 & 7  4 17.79 & -11 28 21.2 &  23   & 14.50 & 13.75 & 13.3 \\
16 & 7  4 45.00 & -11 27 58.9 &  28   & 14.86 & 14.16 & 13.80 \\
17 & 7  4 23.25 & -11 24 17.3 &   5.8 & 13.45 & 12.64 & 12.19 \\
18 & 7  4 16.80 & -11 24 32.4 &  46.2 & 14.12 & 13.21 & 12.60 & [CPD2009] 112\\
19 & 7  4 40.79 & -11 25 37.0 &   7   & 14.38 & 13.57 & 13.26 \\
20 & 7  4 13.53 & -11 24 55.8 &  15.3 & 12.19 & 11.29 & 10.78 & [OSP2002] BRC 27 29\\
21 & 7  4 26.26 & -11 31 20.7 &  17.8 & 12.47 & 11.60 & 11.15 & LkH$\alpha$ 222\\
      \hline
\multicolumn{7}{c}{BRC 45} \\
      \hline
1  & 7 18 36.26 & -22  7 14.6 &   9.7 & 15.71 & 14.87 & 14.67 \\
2  & 7 18 22.93 & -22  4 14.5 &  10.5 & 13.3  & 12.45 & 11.84 \\
3  & 7 18 28.46 & -22  5 15   &  31.5 & 15.07 & 14.03 & 13.36 \\
4  & 7 18 39.85 & -22  0 42.2 &  10.3 & 15.50 & 14.73 & 14.51 \\
5  & 7 19  2.80 & -22  1  4.6 &  38.6 & 13.80 & 12.97 & 12.24 \\
6  & 7 18 56.73 & -22  1  5.3 &  12.8 & 14.89 & 14.09 & 13.55 \\
7  & 7 18 57.67 & -22  1 36.4 &  31.1 & 14.66 & 13.47 & 12.83 \\
      \hline 
    \end{tabular}
  \end{center}
[CDP2009]:\cite{Chauhan09}

[CMP2012]:\cite{Caramazza12}

[DM99]:\cite{Dolan99} 

[BNM2013]:\cite{Bell13} 

[MSB2007]:\cite{Maheswar07} 

[OSP2002]:\cite{Ogura02} 

[PSM2011]:\cite{Prisinzano11} 

[WFT2009]:\cite{Wang09} 

[WTF2008]:\cite{Wang08}
\end{table}

\begin{figure}
  \begin{center}
   \begin{picture}(150,240)(0,-40)
    \put(-81,-138){\includegraphics{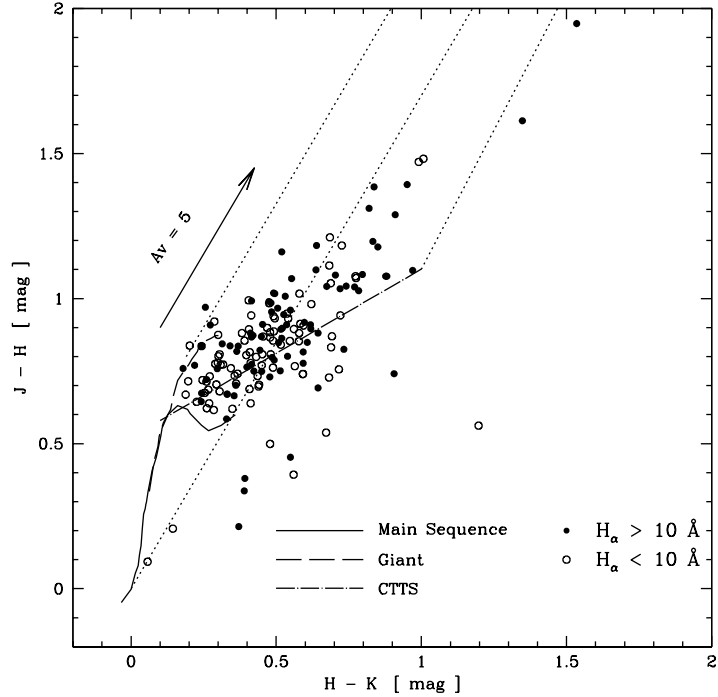}}
   \end{picture}
  \end{center}
  \caption{
	Near-infrared color-color diagram of the H$\alpha$ emission line 
stars in the BRCs.
A certain portion of stars with a weak H$\alpha$ emission line show intrinsic
near-infrared excess.
        }\label{fig:BRCcc}
\end{figure}
}

\sectionn{Discussion}
{ \fontfamily{times}\selectfont
 \noindent
We identified H$\alpha$ emission line stars in the BRCs.
Some objects show intrinsic near-infrared excess
and the others do not.
We investigated the relationship between the intrinsic
near-infrared excess and the H$\alpha$ equivalent widths (Figure \ref{fig:ttstype}).
The intersection of the reddening vector originating from the observed 
$JHK$ colors of the TTSs and the dereddened CTTS line represents the amount 
of the intrinsic near-infrared excess of the TTS.
Zero point of the intrinsic near-infrared excess is defined as the intersection 
of the near-infrared excess border and the dereddened CTTS line.
We also defined the point of unity of the intrinsic near-infrared excess as the 
reddest intrinsic color of CTTSs (Meyer et al. 1997).
The object with the intrinsic near-infrared excess $>$ 0 have an intrinsic 
near-infrared excess.
CTTSs and WTTSs are classified by the H$\alpha$ equivalent width.
We classified the objects into four types.
Type 1 object has the H$\alpha$ emission line with its equivalent width less 
than 10 \AA~ and does not show the intrinsic near-infrared excess.
Type 2 object has the H$\alpha$ emission line with its equivalent width less 
than 10 \AA~ and shows the intrinsic near-infrared excess.
Type 3 object has the H$\alpha$ emission line with its equivalent width larger 
than 10 \AA~ and does not show the intrinsic near-infrared excess.
Type 4 object has the H$\alpha$ emission line with its equivalent width larger 
than 10 \AA~ and shows the intrinsic near-infrared excess.
For the H$\alpha$ emission line stars in the observed BRCs, 30 \% stars are 
classified into
Type 1, 21 \% into Type 2, 25 \% into Type 3, and 24 \% into Type 4.
We do not find mass nor extinction dependencies on the object
types.

\begin{figure}
  \begin{center}
   \begin{picture}(150,230)(0,-40)
    \put(-81,-138){\includegraphics{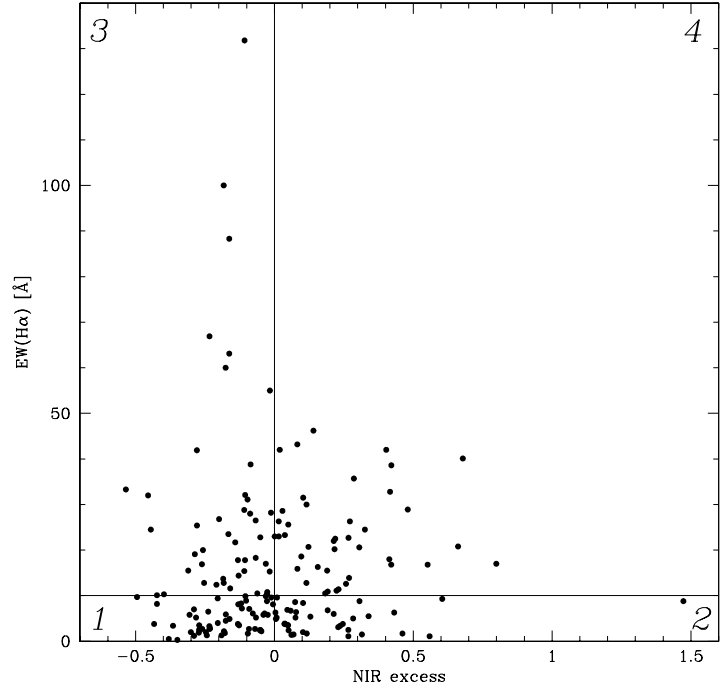}}
   \end{picture}
  \end{center}
  \caption{
	The intrinsic near-infrared excess and the equivalent widths of the 
H$\alpha$ emission line of the TTSs associated with the BRCs.
Type 3 objects and Type 4 objects are identified as CTTSs and Type 1 objects
as WTTSs. Type 2 objects are newly identified with this study.
        }\label{fig:ttstype}
\end{figure}

H$\alpha$ emission line stars in other star forming regions were also 
investigated.
We used the H$\alpha$ equivalent widths listed in \cite{Kenyon95} for the 
objects
in the Taurus molecular cloud, \cite{Frasca15} for the Chamaeleon
molecular cloud, \cite{Erickson11} for the $\rho$ Ophiuchi cloud,
\cite{Ikeda08} and \cite{Nakano12} for BRCs, and \cite{Szegedi13}
for the Orion cluster.
Near-infrared magnitudes of the H$\alpha$ emission line stars
were taken from 2MASS catalog.
Table \ref{tab:type} shows the classification of the H$\alpha$ emission line 
stars in the regions.
It is revealed that Type 2 objects are abundant in the massive
star forming regions such as BRCs compared to
those in the low-mass star forming regions such as the Taurus molecular cloud.
On the other hand, difference in the spatial distributions of Type 2 objects
and the other type objects is not identified for the BRCs observed in this study
and IC 1396 cluster.

\begin{table}
  \caption{Classification of the H$\alpha$ emission line stars}\label{tab:type}
  \begin{center}
    \begin{tabular}{lllll}
      \hline 
Region & Type 1 & Type 2 & Type 3 & Type 4 \\
      \hline 
BRC (this study)       & 52 (30 \%)  & 36 (21 \%) & 44 (25 \%)  & 41 (24 \%) \\
BRC (Ikeda et al.)     & 177 (36 \%) & 108 (22 \%)  & 98 (20 \%)& 108 (22 \%) \\
IC 1396 (Nakano et al.)& 236 (37 \%) & 45 (7 \%) & 249 (39 \%)  & 109 (17 \%) \\
Orion                  & 167 (31 \%) & 16 (3 \%) & 221 (41 \%)  & 135 (25 \%) \\
Taurus                 & 53 (42\%)   & 5 (4 \%) & 25 (20 \%)    & 42 (34 \%) \\
Chamaeleon             & 44 (63 \%)  & 3 (4 \%) & 14 (18 \%)    & 12 (15 \%) \\
$\rho$ Oph             & 185 (81 \%) & 7 (3 \%) & 21 (9 \%)     & 16 (7 \%) \\
      \hline 
    \end{tabular}
  \end{center}
\end{table}

The general concept that a low-mass star evolves from a CTTS to a WTTS
is well established by many observational studies of nearby
low-mass star forming regions.
In our definition, Type 3 and Type 4 objects correspond to CTTSs and
Type 1 objects to WTTSs.
Dissipation process of a circumstellar disk around a low-mass star
in strong UV field emanating from a nearby OB star has been widely discussed.
\cite{Stolte10} observed the Arches cluster near the Galactic center.
They identified a significant population of near-infrared excess sources.
The disk fraction of B-type star was derived as 6 \% in the Arches
cluster.
On the other hand, the fraction was as low as 3\% in the
vicinity of O-type stars in the cluster core.
They concluded that disk dissipation process was more rapid
in compact starburst clusters than in moderate star-forming environments.

Disk dissipation process due to UV radiation
is also examined by numerical simulations.
\cite{Anderson13} considered circumstellar disk evolution in strong far-UV
radiation fields from external stars.
It is revealed that the UV radiation from nearby OB stars heats the gas
near the disk edge and effectively drives mass loss from circumstellar
disks.
They also found that the UV radiation photoevaporates disks and disk 
radii are truncated to less than $\sim$ 100 AU.

Type 2 objects have a weak H$\alpha$ emission line but
show the intrinsic near-infrared excess.
Those are classified into WTTSs from optical spectroscopy, albeit
into CTTSs from near-infrared photometry.
The general concept of the formation process of a low-mass stars does not
involve such objects.
We propose two hypotheses for Type 2 objects.
The first hypothesis involves the idea that a Type 2 object is a WTTS with 
a flaring circumstellar disk.
The H$\alpha$ emission line stars are associated with the 
BRCs. Photons from the nearby OB star ionize hydrogen atoms outside the BRCs
and excite hydrogen atoms at the boundary of the BRCs.
Type 2 objects may be irradiated by UV photons from the nearby OB star.
\cite{Walsh14} calculated the structure of a circumstellar disk irradiated
by UV radiation emanating from a nearby massive star, 
based on the circumstellar disk model of \cite{Nomura05} and \cite{Nomura06}.
They indicated that the disk has large scale height,
because gas in the disk is heated by the UV radiation then expands.
We constructed SEDs of the H$\alpha$ emission line stars
in IC 1396 cluster with the photometric data of Guide Star Catalog,
2MASS catalog, and WISE catalog, then fitted them with the SED model of 
\cite{Robitaille06}.
Surface height of the circumstellar disk is expressed as,
\begin{equation}
H(r) = H_{0}(\frac{r}{R_{0}})^\beta,
\end{equation}
where $H_{0}$ is the disk half-thickness, $R_{0}$ is the radius of the
central star, and $\beta$ is the flaring parameter \cite{Kenyon87}.
$\beta$ is deduced to be between 1.00 and 1.20 for
all objects.
The average $\beta$ are $1.067\pm0.008$, $1.095\pm0.015$, and $1.106\pm0.009$
for Type 1, 3, and 4 objects, respectively,
while that of Type 2 objects is as high as $1.125\pm0.011$.
Large $\beta$ of Type 2 objects supports the idea that those are 
WTTSs with a highly flared
disk, although the difference in $\beta$ between Type 2 and Type 4 is
less than 2 $\sigma$ significance level.
Spatial distribution of Type 2 objects should be inhomogeneous,
if this hypothesis is valid. One may guess rich population of Type 2
objects outside the BRC or at the surface of the BRC.
However, we do not find such spatial distribution of Type 2 objects.

Another hypothesis is that Type 2 objects are pre-transitional disk objects.
A transitional disk object has a circumstellar disk with an inner hole
created by photo evaporation of a central star or planet
formation.
A pre-transitional disk object has a small and
optically thick disk in the 
inner hole of the circumstellar disk.
If near-infrared excess arises from the inner disk and a small amount of 
material accretes from the inner disk to the photosphere, then the object 
is classified into Type 2 object. 
We investigated H$\alpha$ strength and near-infrared excess of five 
pre-transitional disk objects listed in \cite{Marel16}.
Four objects are located in the Type 2 region, or
in Type 4 region near the border between Type 2 region and Type 4 region
in the near-infrared excess and H$\alpha$ equivalent width diagram.
Similarities between Type 2 objects and pre-transitional disk objects is 
also found in the near-infrared and WISE colors.
We plotted the TTSs
in the BRCs, the transitional disk objects in the L 1641 cloud \cite{Fang13},
and the known pre-transitional disk objects on the 
near-infrared and WISE color diagram (Figure \ref{fig:wisecc}).
Majority of Type 2 objects are not plotted in the 
transitional disk object region, but their $K -$[4.6] colors are
redder than that of the transitional disk objects.
The known pre-transitional disk objects and Type 2 objects in the BRCs are
plotted on the similar region in the diagram.

\begin{figure}
  \begin{center}
   \begin{picture}(150,230)(0,-40)
    \put(-81,-138){\includegraphics{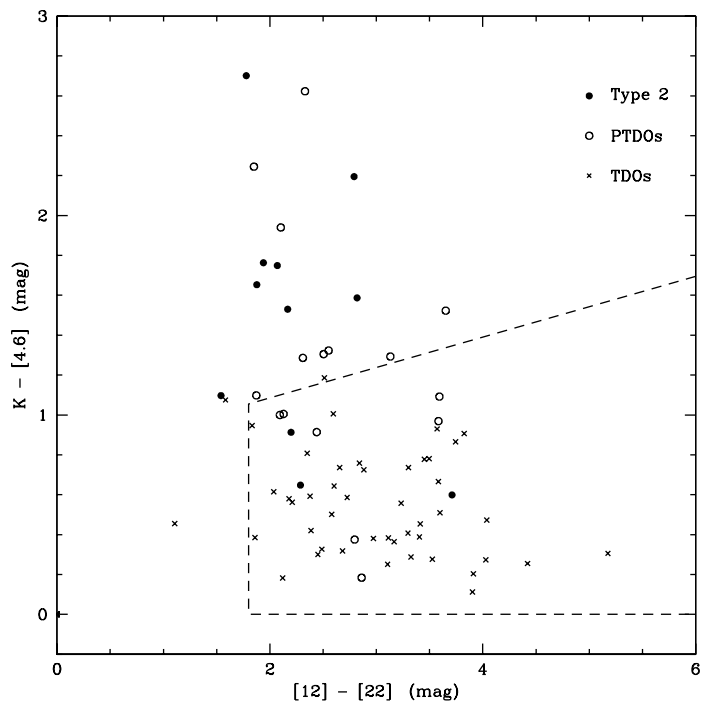}}
   \end{picture}
  \end{center}
  \caption{
	The near-infrared $K$-band and WISE 4.6$\mu$m-, 12$\mu$m-, and 
22$\mu$m-bands color diagram of Type 2 objects, transitional
disk objects, and pre-transitional disk objects.
The region enclosed by a dashed line is the region of the transitional
disk objects.
Type 2 objects and pre-transitional disk objects have redder color in $K$ - 
[4.6] color.
        }\label{fig:wisecc}
\end{figure}

\cite{Kusune15} carried out near-infrared polarimetry of BRC 74.
They found that the magnetic field in the layer just behind the rim ran
along the rim.
The estimated magnetic field strength was $\sim 90 \mu$G,
stronger than that far inside, $\sim 30 \mu$G,
thereby suggesting that the magnetic field inside the rim is 
enhanced by the UV-radiation-induced shock.
A proto-planetary disk has an inner
stable region and an outer unstable region, if it magnetizes.
\cite{Sano00} indicated that the boundary between stable and unstable regions
is located at $\sim$ 20 AU from the central star and the stable region shrinks
in strong magnetic field environment.
An abundant population of Type 2 objects in the BRCs may indicate slow evolution
from CTTSs to WTTSs.
Geometric structure and evolution timescale
of a proto-planetary disk in the close vicinity of the 
central star under a strong magnetic field is to be investigated.
}

\sectionn{Conclusions}

{ \fontfamily{times}\selectfont
 \noindent
We have conducted slit-less optical spectroscopy for 14 bright rimmed clouds
and found 173 H$\alpha$ emission line stars.
Among them, 36 objects have a weak H$\alpha$ emission line, but show
intrinsic near-infrared excess.
Those are identified as WTTSs with optical spectroscopy, but
as CTTSs with near-infrared photometry.
The general concept of the formation process of a low-mass star
does not involve such objects.
Those might be weak-line T Tauri stars with a flared circumstellar disk
in which gas is heated by ultraviolet radiation from a nearby early-type star.
Alternatively, those might be pre-transitional disk objects.
}

 {\color{myaqua}

 \vskip 6mm

 \noindent\Large\bf Acknowledgments}

 \vskip 3mm

{ \fontfamily{times}\selectfont
 \noindent
We thank the telescope staff members and operators at the IUCAA 2m Telescope.
This study was partly supported by the JSPS-DST collaboration.

 {\color{myaqua}

}}

\end{document}